\documentstyle[12pt]{article}
\begin{document}
\title{Relevance of Dynamic Clustering to Biological Networks}
\author{
        Kunihiko KANEKO \\
        {\small \sl Department of Pure and Applied Sciences}\\
        {\small \sl University of Tokyo, Komaba, Meguro-ku, Tokyo 153, JAPAN}
\\}
\date{}
\maketitle
\begin{abstract}
Network of nonlinear dynamical elements often show
clustering of synchronization by chaotic instability.
Relevance of the clustering to ecological, immune, neural,
and cellular networks is discussed, with the emphasis of partially
ordered states with chaotic itinerancy.
First, clustering with bit structures
in a hypercubic lattice is studied.  Spontaneous formation
and destruction of relevant bits are found,  which give self-organizing,
and chaotic genetic algorithms.  When spontaneous changes
of effective couplings are introduced, chaotic itinerancy of clusterings is
widely seen through a feedback mechanism, which supports dynamic
stability allowing for complexity and diversity, known as homeochaos.
Second, synaptic dynamics of couplings is studied in relation with
neural dynamics.  The clustering structure is formed with a balance between
external inputs and internal dynamics.  Last, an extension allowing for
the growth of the number of elements is given, in connection
with cell differentiation.  Effective time sharing system of resources
is formed in partially ordered states.
\end{abstract}
\section{Dynamical Viewpoints}
As is discussed in the preface,
dynamical viewpoints have been appreciated in biological sciences.
Here we need some logic to understand complex dynamical networks.
Such studies are required from neural, immune, cellular,
and ecological networks.
In neural systems, Tsuda has stressed
the importance of (chaotic) dynamics in functions
over several years \cite{Tsuda-book,Tsuda0,Tsuda-CI,Tsuda00}.
Freeman has noticed the importance of the
change of the degree of coherence of neural activities \cite{Freeman}.
In the epilepsy, an ensemble of neurons exhibits
a large spike due to the coherent oscillation of
neural activities. Partial synchronization of nonlinear oscillations has been
discovered in the visual cortex of a cat \cite{Eckhorn,Singer}.
Vaadia and Aertsen \cite{Kruger-book,Aertsen}
have found that the effective coupling among
neurons varies temporally in a rather short time scale.
They have found that the degree of synchronization of
oscillations change both temporally and by the choice of pairs.
In an ecological system, many species coexist in a network of food web.
The population dynamics of species seems to be more stable
as the complexity of network is larger,
as Elton has discovered in the forest of England \cite{Elton}.
Furthermore the stability
may not be sustained as a fixed point state \cite{May}, but is
sustained in a dynamically changing state.
In tropical rain forests, for example, there are a variety of
species each of which has small population.
Temporal variation of populations there is so large that
the diversity in rain forests is often believed to be maintained only
in a nonequilibrium state \cite{rainforest}.
Similar interacting population dynamics is also important in
the immune network, where Jerne proposed the network of
antigens and antibodies \cite{Jerne}.  Possibility of many
attractors in such network system is discussed \cite{Pa-Per} in relation
with spin glass type models \cite{SG}, while
temporally successive switches of many states are
discussed in \cite{Ike}.
At a somatic level, metabolic reactions often show nonlinear oscillations
through catalytic reactions.
In a developmental process and cell differentiation,
interaction among cells is important
besides the control by gene switches.
Creation of diverse cells by the latter mechanism
is often discussed in relation with
many fixed point attractors \cite{Kauff}, while dynamical viewpoints in
cellular interactions are stressed in a recent experiment\cite{Yomo,KKTY}.
In these fields, studies on dynamic nonlinear networks are
strongly requested, where a huge number of
interacting nonlinear elements is involved.
So far, spin-glass type models are used as a standard one
for a system with many fixed-point attractors organized
as a tree structure.  With such models, static aspects
or relaxation towards stationary states are studied.
To address the dynamical problems listed above, however, we need
studies of a system with many nonlinear interacting elements.
The purpose of the present paper is to point out that
the network of chaotic elements can provide a novel
standard framework for a variety of biological networks
with dynamical complexity.
The thesis of the present paper is motivated by
the previous studies by the author on an ensemble of chaotic elements.
There \cite{KK-GCM,GCM2} it was found that clustering of synchronization is
a general and important feature in globally coupled dynamical systems.
Elements split into few or many clusters, in which
their oscillations are synchronized.
The number of clusters can differ by attractors, and by the strength of chaos.
Complex partition into clusters is also found. This complexity is
partly common with the spin glass type problems \cite{GCM-part}.
Generally speaking, there are three possibilities in clustering;
phase, amplitude, and frequency of oscillations.  For example,
in the pure phase clustering, the amplitudes and periods of oscillations
of elements are identical; only the phases of oscillations differ by
clusters to which elements belong.  So far the clustering we have studied
does not purely consist of only one of the above three types.
Phase, amplitude, and frequency clusterings are mixed, although
 the phase difference is most relevant to clusterings.
In neural systems, oscillations seem to split into clusters, as
discussed as "gravitational clustering" by Vaadia and Aertsen \cite{Aertsen}.
These clusterings are rather complicated, although
the phase differences seem to be most important.
In clustering it should be noted that identical chaotic elements differentiate
spontaneously into different groups:
Even if a system consists of identical elements,
they split into groups with different phases of oscillations.
Hence a network of chaotic elements gives
 a theoretical basis for differentiation of
identical elements, and provides a mechanism on the
origin of diversity and complexity in biological networks.
Besides this complexity in fixed relationships ( since
two elements in the same cluster remain at the same cluster there),
dynamical changes of relationships and synchronizations
are of importance in the biological problems listed above. Indeed,
the network of chaotic elements  shows
dynamical complexity, when chaotic instability in each element is stronger;
a typical example here is chaotic
itinerancy \cite{KK-GCM,GCM-CC,Tsuda-CI,Ikeda}(see also \S 2).
The maintenance of diversity and complexity,
besides their origin, is also an important problem in an evolutionary
system.  A dynamical mechanism of maintenance of diversity
is recently proposed as homeochaos \cite{homeo,homeo2}.  Here we also discuss
a possible relationship between homeochaos and clustering.
The purpose of the present paper is  to survey the relevance of
the idea of clustering to biological systems.  For the application
of the idea, it is often necessary to extend the basic network of chaotic
elements to different topology, to non-uniform elements, to
synaptic couplings, and to systems with variable degrees of freedom.
In the next section we give a brief review of basic results of clustering,
and chaotic itinerancy in globally coupled maps (GCM).
In \S 3, the clustering idea is extended to a system with hypercubic topology.
This extension is motivated by interacting population dynamics
with mutation and its application to genetic algorithms.
The formation of synchronized clusters strongly reflects
the bit structure in the lattice.
Indeed we will see self-organization and destruction of relevant bits
and ``don't care" bits by the chaotic itinerancy mechanism.
In \S 4, we further extend the system in \S 3 to
allow for a change of the coupling strength.  The motivation for
this extension comes from a system with interacting populations
with mutation of mutation rates.  By the
last process, the coupling strength among elements (species)
is effectively changed with time.
It turns out that the system attains a dynamic stability
allowing for diversity of many groups, by
forming a feedback mechanism to adjust the coupling strength.
This mechanism, called {\sl homeochaos} turns out to be
sustained by successive changes of clusterings.
\S 5 and \S 6 are devoted to extensions of our GCM to synaptic coupling
cases, motivated by applications to neural systems.
In \S 5, a globally coupled map with distributed coupling strengths
is shown.  Different types of clustering behaviors,
from synchronized to completely desynchronized, are observed
within a unique system.  This observation opens up the
possibility of controlling the degree of synchronization of elements
 according to inputs,  by modifying the coupling strengths among the elements.
Such control is carried out by a synaptic model introduced in \S 6,
which is in a possible relationship with
the synchronization by external inputs in the brain \cite{Kruger-book}.
In \S 7, we study clusterings in a system with growing degrees of freedom,
in connection with cell differentiation and growth.  It is found that the
dynamic clustering leads to growth of the number of cells
by forming a time sharing system of foods
(resources).  \S 8 is devoted to a brief summary and discussions.
A rather personal view on biological
networks from the above lines is given in Table I.
\vspace{.2in}
\begin{tabular}{|l|l|l|l|l|}\hline
Field & One-to-one & Static  & Dynamic& Key   \\
       & map       &  Complex & Complex & Concepts  \\ \hline
Neuroscience & Grand-  & Typical & Dynamical & Clustering \\
        & mother Cell       & neural net   & Correlation
\cite{Tsuda00,Tsuda-book,Kruger-book} & CI \\ \hline
Ecology & Niche- & Random & Dynamic ecological&Clustering\\
     & -species & network \cite{May} & network\cite{homeo}& Homeochaos \\
\hline
Immune & antigen- & Jerne's   & Jerne's net & \\
        & antibody & net and \cite{Pa-Per} & net and \cite{Ike}& Homeochaos? \\
\hline
Development & gene-enzyme  & Kauffman's   &  GCM type& Clustering\\
        &   etc.    & net\cite{Kauff} & \cite{KKTY}& Open Chaos \\ \hline
Basic Model & ---& Spin-glass & GCM  & ---\\
  & & type\cite{SG}& type \cite{KK-GCM}& \\ \hline
\end{tabular}
Table I   Views of biological networks ( very personal perspective)
\hspace{3in}CI=Chaotic Itinerancy
\vspace{.2in}
\section{Brief Review of Globally Coupled Maps}
The simpliest case of global interaction is studied
as the ``globally coupled map" (GCM) of chaotic elements \cite{KK-GCM,GCM2}.
An example is given by

\begin{equation}
x_{n+1}(i)=(1-\epsilon )f(x_{n}(i))+\frac{\epsilon }{N}\sum_{j=1}^N f(x_{n}(j))
\end{equation}
where $n$ is a discrete time step and $i$ is the index of an
element ($i= 1,2, \cdots ,N$ = system size), and $f(x)=1-ax^{2}$.
The model is a mean-field-theory-type extension of coupled map
lattices (CML) \cite{CML}.  The above dynamics consists of parallel nonlinear
transformation and a feedback from the ``mean-field".  It is
equivalent to
\begin{math}
y_{n+1}(i)=f[(1-\epsilon )y_{n}(i)+\frac{\epsilon }{N}\sum_{j=1}^N y_{n}(j)],
\end{math}
with the aids of transformation $y_n (i)=f(x_n (i))$.
In this form, one can see clear correspondence with
neural nets:
If one chooses a sigmoid function (e.g.,$tanh(\beta x)$)
as $f(x)$ and a random or coded coupling $\epsilon _{i,j}$,
a typical neural net is obtained.
Through the interaction, some elements oscillate synchronously,
while chaotic instability gives a tendency of destruction of the coherence.
Attractors in GCM are classified by the number of synchronized
clusters $k$ and the number of elements for each cluster $N_k $.
Here a cluster is defined
as the set of elements in which $x(i)=x(j)$ \cite{Nakagawa}.
Each attractor is coded by the
clustering condition $[k,(N_1 ,N_2 ,\cdots,N_k )]$.
If we distinguish each element $i$,
there are $N!/(N_1 ! N_2 ! \cdots N_k !)$ ways of the
partitions for each clustering condition $(N_1 ,N_2 ,\cdots
,N_k )$. We have exponentially many attractors per
each clustering condition.
An interesting possibility in the clustering is
that it provides a source for diversity.  Even if the system is
started from identical states, they split into different groups.
In \S 7, we discuss the possibility
of a role of clustering in cell differentiation.
In a globally coupled chaotic system in general,
the following phases appear successively with the increase of
nonlinearity in the system ($a$ in the above logistic map case)
\cite{KK-GCM}:
(i) {\bf Coherent phase}: A coherent attractor ($k=1$)
has occupied (almost) all basin volumes.

(ii) {\bf Ordered phase}: Attractors ($k=o(N)$) with few clusters
have occupied (almost) all basin volumes.

(iii) {\bf Partially ordered phase}: Coexistence of attractors
with many clusters ($k=O(N)$) and few clusters.

(iv) {\bf Turbulent phase}:  All attractors have many ($k=O(N)$;
in most cases $k \approx N$) clusters.

In the turbulent phase,
although $x(i)$ takes almost random values almost independently,
there remains some coherence among elements.
Indeed the distribution of
the mean field $h \equiv (1/N)\sum _j f(x(j))$ does not
obey the  law of large numbers.  The emergence of hidden
coherence is a general property in a globally coupled
chaotic system \cite{GCM2}.
Existence of such coherence may be important to discuss about the EEG.
In EEG, one measures a given average of neuronal (electric) activities.
Since a firing pattern of each neuron is not regular
( i.e., chaotic or random),
the amplitude of the variation of EEG might decrease with the
number of neurons involved in the average.  Still, we have observed a
large enough amplitude of variation in EEG.  This observation suggests
that there remains some correlation among neuronal bursts.
If each neuronal bursting were random, it would be hard to imagine a mechanism
to keep such coherency.  The above hidden coherence in globally
coupled chaotic systems may be the origin of such coherence.
In the partially ordered (PO) phase,
complexity of partition into clusters is high.
There are a variety of attractors with a different number of clusters,
and a different way of partitions $[N_1,N_2, \cdots ,N_k]$.
We have measured the fluctuation of the partitions,
using the probability $Y$ that two elements fall on the same cluster.
This $Y$ value fluctuates by initial conditions.  This fluctuation seems
to remain finite even if the size goes to infinity \cite{GCM-part}.
Such remnance of fluctuation of partitions is also noted in
spin glass models \cite{SG}.
In the PO phase, this fluctuation is enhanced.
This increase of complexity
at the partially ordered state may be important to
study the relevance of the PO phase to biological systems.
We also note that the partition is usually inhomogeneous, and
is organized as an inhomogeneous tree structure as in the spin glass
model \cite{KK-GCM,SG}.
Besides the above {\sl static} complexity, there emerges dynamic complexity
in our model at the PO phase.  The orbits make itinerance over
ordered states via highly chaotic states.  In the ordered
states the motion is partially coherent.
Our system exhibits {\bf intermittent changes between
the self-organization towards the coherent
structure and its collapse to a high-dimensional
disordered motion}.  This dynamics, called {\sl chaotic itinerancy},
has been found in a model of neural dynamics by Tsuda \cite{Tsuda-CI},
optical turbulence \cite{Ikeda}, and in GCM.
Here a number of ruins of low-dimensional attractors coexist in the
phase space.  The total dynamics consists of the residence at a ruin and
a high-dimensional chaotic state interspersed between the residence.
\section{Bit Clustering in Hypercubic Coupled Maps:
Basis of Chaotic Genetic Algorithm}
Let us discuss some (population) dynamics of many individuals,
coded by genes.
If genes are represented by a bit sequence, the mutation process
in gene space is given by the diffusion in the bit sequence.
When some nonlinear population
dynamics is included to take into account of
the saturation, competition, or prey-predator (host-parasite)
interaction, the total population dynamics
is given by the local nonlinear dynamics and the diffusion process on the
hypercubic lattice of length 2, corresponding to the bit
sequences.   The minimal model for this process is given by
the following coupled map on a hypercubic lattice;
\begin{equation}
x_{n+1}(i)=  (1-\epsilon )f (x_n (i))+ \frac{\epsilon}{K} \sum_{j=1}^K
f(x_n(\sigma _j(i))),
\end{equation}
where $\sigma _j (i)$ is a ``species" whose $j$'th bit is different from
the specie $i$ (with only one bit difference) ,
and $K$ is the total bit length of species ( the number of
total ``species" is $2^K$).
We use a decimal representation of bit sequence often;
for example 42 stands for the sequence 101010, and $\sigma_2(42)=40$
The present model may be relevant to genetic algorithms \cite{GA},
where population of bit-strings changes according to their fitness.
The model is also of theoretical interest, since it lies between
globally coupled and locally coupled models (CML) \cite{CML}:
In a global coupled chaotic system, we have $N$ connections per element,
while a d-dimensional CML (with nearest neigbor coupling) has $2d=o(N)$
connections per elements.  In our hypercubic system with  $N=2^K$ elements,
we have $K=log_2 N$ connections per elements \cite{Chate}.
In the model (2) we have often observed a state with few synchronized clusters
when the nonlinearity parameter $a$ is not large.
Here a synchronized cluster means, as in \S 2, that  two elements
in the cluster oscillate in complete synchronization, i.e.,
$x_n(i)=x_n(j)$ for two elements $i$ and $j$ in the cluster.
Here the split to two clusters is organized according to the
hypercubic structure.  For examples, following types of clusterings are
observed.
(A) 2 clusters by 1 bit:
Elements split into two synchronized clusters.  All elements in
each cluster oscillate in synchronization as shown in Fig.1a) and b).
In Fig. 1, successive snapshots of $x_n(i)$ are plotted as a function of $i$.
Such clustering into two is frequently observed in a
globally coupled map system, where the partition into two clusters is
arbitrary. In our HCM, the partition
is governed by the spatial structure in the hypercubic lattice.
For example, elements may be grouped into two clusters
with **0*** and **1***, (* means that the symbol there is either one or zero),
each of which has $2^{K-1}$ species.
If the elements  split into the group of ******1 and ******0 for example,
the snapshot of $x(i)$ shows a zigzag
structure ( with period 2), if plotted as a function of the
decimal representation $i$ (see Fig.1a),
while the split by the K-th bit leads to a periodic
structure with period $2^{K}$.
This clustering is formed by cutting the
K-dimensional hypercube by a hyperplane (see Fig.1b).
(B) 2 bit clustering
Depending on initial conditions and parameters, the number of
relevant bits for clustering can be larger than the case (A),
as well as the number of clusters.
In the 2 bit clustering there are following possibilities
(B0) 4-cluster state with two relevant bits:
This case is just a direct product of the previous case (A).
The hypercubic space splits by two hyperplanes.
Elements split into
four clusters, for example, coded by 01*****, 10*****,
11*****, and 00*****.
(B1) 2-cluster with 2 bits (XOR construction)
We have also found an attractor with 2 clusters with the use of
2 relevant bits.  For example the elements split into
the groups (i) 10***** or 01***** and (ii) 00***** or 11*****.
This split corresponds to the construction of XOR (exclusive or)
with the use of 2 relevant bits (see FIg.1c).
(B2) 3-cluster with 2 bits
We have also found a 3-cluster state with 2 bits, constructed
for example as (i)10***** or 01***** (ii) 11***** and (iii) 00*****.
Here we have to note that not all partitions are
possible in the HCM.  Even if we start from
an initial condition with a given clustering condition,
the synchronization condition ($x(i)=x(j)$
for $i,j$ belonging to a same cluster)
is not satisfied at the next step, for
most of such initial conditions.
In contrast with the GCM case, not all possible partitions
can be a ( stable or unstable) solution of the evolution
equation.
One can easily check that the synchronization is
preserved for the clustering (A), (B0), (B1), and (B2).
Generally the clustering should be constructed
as a combination of hyperplane cuts, and the
condition of a cluster is written as a bit representation
with the symbol ``*", corresponding to the ``don't care" bit ``\#",
in genetic algorithms \cite{GA}.
(C) 3 bit and higher ($K$) bit coding:
Clustering with the use of bits more than 2 are
constructed in a similar manner.  Most clusterings observed
here are direct products of (A) or (B1)-(3).
(C1) parity check:  2 clusters from $K$ bits;
elements split into two groups according to the parity of the number of
1's in each bt representation.  For example, elements
split into two clusters as follows:
(i) 000, 011, 101, 110 and (ii) 001 010, 100, 111, for $K=3$.
The clustering, thus gives a parity check.
It is a hypercubic version of the  zigzag (1-dim)  or checkerboard (2-dim)
pattern \cite{CML}.
(C2) Hamming distance code : $K+1$ clusters from  $K$ bits;
elements split into $K+1$ clusters according to the Hamming
distance from an element. This is a straightforward extension of
the clustering (B3).  For example,
 elements split into 4 clusters as follows
(i) 000 (1 element) , (ii) 001 010, 100 (3 elements) , (iii)
011,101, 110, (3 elements) and (iv) 111 (1 element) , for $K=3$.
This clustering is constructed by the cuts by $K$ parallel hyperplanes.
An attractor with many clusters is often constructed
by a direct product of combinations of (A0), (C1), and (C2) ,
that is of (0) 1 bit code by a hyperplane cut
(1) Parity check and (2) Hamming distance cuts.
( see for example Fig. 1d) for a 6-cluster attractor).
If local chaos is stronger ($a$ is larger), the clustering
is not complete:  Some elements stay very close and oscillate almost
synchronously over some time steps, but then they are separated due
to chaotic instability.  At some medium nonlinearity regime,
we have found a chaotic itinerancy
state, where relevant bits change according to temporal evolution.
In Fig.2, change of relevant bits for the clustering is seen.
So far this state is seen only in a very narrow parameter regime,
and is found as very long transients before the system finally
falls on an attractor with few clusters.
With the introduction of  external inputs to each element, it is also possible
to have a clustered state following the external information
\cite{HCM}.  Relevant information is extracted through this process
spontaneously, which is stored as a relevant bit in the clustering.
\vspace{.2in}
Fig. 1
\vspace{.1in}
Fig.2
\section{Homeochaos and Clustering:  Dynamic Maintenance of Diversity}
In a system with interacting population dynamics,
it is an interesting question how diversity of
genes are maintained.
In \cite{homeo}, we have proposed a concept ``homeochaos" as a
mechanism for sustaining dynamic stability with diversity.
There, population dynamics models
with interaction among species, mutation, and mutation of
mutation rates \cite{homeo,homeo2} has been studied.
In particular, we take a 2-dimensional map for local dynamics here,
since we are interested in the interaction between
hosts and parasites (or preys and predators).
Each individual has a gene coded by a bit sequence as in \S 3.
We assume that the parasite can attack a host only if their
bit strings are completely matched.  By this restriction, we
can have a 2-dimensional map for the local population dynamics of
each bit string.  To be specific we have studied the model
\begin{equation}
h'(i)=a (1- h(i)) (h(i)) exp(-\beta p(i))
\end{equation}
\begin{equation}
p'(i)= h(i)(1-exp(-\beta p(i)))
\end{equation}
instead of the 1 dimensional logistic map in \S 3.
Here the set of variable $(h(i),p(i))$ gives the population of the
host and parasite of the ``species" $i$.
The term $exp(-\beta p(i))$ represents the fraction that is
killed by the corresponding parasite $i$.
Further we assume that each ``species" $i$ can have different mutation rates,
in other words, each is coded by $(i,j)$ rather than $i$.
Each group $(i,j)$ has a mutation change from its nearest neighbor point
in the hypercubic lattice $\sigma(i,k)$ in \S 3 with the rate
given by the mutation level $j$
(here we assume $\epsilon _j=2^{(j_{max}-j)/4}$ and $j_{max}=30$).
Since the mutation level does not affect the
interaction, the local dynamics is obtained straightforwardly from (3) and (4);
First introduce  $h^s (i)=\sum_j h_n(i,j)$ and $p^s(i)= \sum_j p_n (i,j))$,
the sum of the population of each ``species" over all mutation levels.
Then the local dynamics of $h^s (i)$ and $p^s(i)$ obey exactly (3) and (4).
The local dynamics $h(i,j) \rightarrow h'(i,j)$
is given by multiplying the dynamics of $h^s (i)$ by $h_{n}(i,j) /h^s_{n}(i)$,
the fraction of the population of the level $j$.
( Of course the dynamics $p(i,j) \rightarrow p'(i,j)$ is given by
multiplying the corresponding equation by $p_{n}(i,j) /p^s_{n}(i)$.)
Thus the total dynamics, (which is a little bit complicated but is
straightforwardly obtained ) is given by the coupled map lattice
with the diffusive coupling by the mutation.  The coupling strength
$\epsilon _j$ depends on the level $j$, as stated, and there is also
diffusive coupling
between neighboring mutation levels $j \rightarrow j \pm 1$, with
the (mutation ) rate $\epsilon _j $ ( see \cite{homeo2}).
%
Through the mutation of mutation rate, the mutation level is sustained
at a high level, with some temporal fluctuation. As is studied in
\cite{homeo2},
weak high-dimensional chaos is found in this case, which
affords the stability with diversity.  Such stability,
called ``homeochaos" is proposed to be essential to the stability of
biological networks \cite{homeo,Alife}.
In Fig.3, we have plotted snapshots
of $h^s(i)$ successively, where
changes of bit clustering are found with
chaotic itinerancy.  In Fig.4a),
we have also plotted the time series of the effective degrees of freedom,
defined as the number of clusters with a given (finite) precision;
i.e., $h^s(i)$ and $h^s(j)$ agree within the precision, they are assumed
to be the same ``effective" cluster.  Drops of the effective
degrees give the emergence of ``almost" clustered
states as in Fig.3a).  Corresponding time series of the average
mutation level is given in Fig.4b).
This type of chaotic itinerancy is seen in a narrow parameter regime
in the previous section ( e.g. only around $a\approx 1.52$ in the
model in \S 3 for $\epsilon \approx .3$), as long transients.
In the above two variable model, it is seen only
around $a \approx 3.75$ (as long transients),
when the mutation rate is fixed ( e.g., at $j=25$).
On the other hand, such behavior is seen over all regimes
for $a>3.5$, with the inclusion of mutation of mutation rates.
Hence  there must be a mechanism
to adjust the effective coupling strength so that
( i.e., the mutation rate here) the system stays around at
the partially ordered state.  Such adjustment mechanism
is important in homeochaos.
A high mutation level is necessary to have a state with few clusters.
As the mutation level is increased, the oscillation tends to be synchronized
 ( recall that the number of
clusters in coupled maps decreases with the increase of coupling strength;
see \S 2 and \S 3).
By examining the change of mutation rates (coupling strengths)
and the change of clustering in detail, we have
found the following feedback mechanism to keep the system
at the partially ordered state:
(i) increase of mutation level leads to the synchronization of population
oscillation, since the coupling is increased :
(ii) the synchronization leads to decrease of the mutation level \cite{homeo2}:
(iii) the decrease of mutation level leads to the split of synchronized
clusters, since the coupling is decreased ( see \S 2 and \S 3).
(iv) split of clusters is associated with the irregular temporal dynamics,
which leads to increase of mutation level, and then to (i) \cite{homeo2}.
This feedback mechanism is necessary to maintain our system at
the partially ordered state with the chaotic itinerancy of bit clusterings.
Since such partially synchronized state is thought to be
important in biological networks \cite{Tsuda-CI,Aertsen},
the above mechanism for evolving to, and maintaining,
the state is important.  It is important to search for a similar
feedback mechanism in other biological networks.
\vspace{.3in}
Fig.3  Snapshot showing clustering
Fig.4  a) effective degrees of freedom b) mutation level
\section{Extensions to Coupling forms with more complicated structure}
In a biological system, the elements are not homogeneous.
It is often necessary, then, to assume that local dynamics or a
coupling strength depends on elements.  Also time dependence of
such parameters may be necessary.
Here we extend the previous GCM by allowing for inhomogeneous
parameters by elements, instead of identical parameters for all.
There are two possibilities here; choice of distributed $a(i)$ instead of
a constant nonlinearity $a$ or distributed $\epsilon (i)$ instead of a
constant coupling $\epsilon$.  For the latter case, the model
is given by
\begin{equation}
x_{n+1}(i)=(1-\epsilon (i))f(x_{n}(i))+\frac{\epsilon (i)}{N}\sum_{j=1}^N
f(x_{n
}(j)).
\end{equation}
In Fig.5 we have plotted a snapshot of $x_n (i)$ for a system with
homogeneously distributed coupling over
$[\epsilon _{min}, \epsilon_{max}]$ (i.e.,
$\epsilon(i)=\epsilon_{min} + (\epsilon_{max} -\epsilon_{min})\frac{i}{N}$).
As is seen, the clustering structure depends on the coupling strength
at each element.  Elements with large coupling ( $\epsilon> .22$) form
a single synchronized cluster, while the number of clusters increases
successively with the decrease of the coupling strength at the element.
Such clustering bifurcation looks similar with the phase change i)
$\rightarrow$ iv) in \S 2.  However, this is not a trivial extension.
The clustering change in \S 2 is the bifurcation with
system parameters, while the change here is included in a
a single network system, where
all the elements therein are connected by a unique mean field.
Still the internal bifurcation among elements occurs here.
At the edge parameter region between clustered and turbulent states
the motion is rather complicated with chaotic itinerancy.
Desynchronized bursts emitted from the elements with smaller $\epsilon (i)$
flow to elements with larger $\epsilon (i)$, where clustering
can change in time.
The behavior discussed here is also seen in a model with
the distributed nonlinearity, i.e.,
$a(i)=a_{min} + (a_{max} -a_{min})\frac{i}{N}$.
If the range of distribution of coupling is small, the behavior there
approaches that in the previous section.  According to the width of
the coupling range variety of clusterings decreases.  We note that the
hidden coherence still emerges in the turbulent state, even if the
parameters $a$ or $\epsilon$ are distributed \cite{GCM2}.
\vspace{.2in}
Fig.5  x(i) vs i  a) Snapshot b)Overlaid
\section{Synaptic Dynamics}
In traditional neural networks, coding is assigned at firing patterns
$x(i)$, while the synaptic change and memory are assumed to be
assigned on the interaction strength between two elements $i$ and $j$.
Recently there are some arguments that
dynamical coding, given by
the correlation between $x(i)$ and $x(j)$, may be important
\cite{Kruger-book,Aertsen}.
Then it may be interesting
to pursue this ``converse" limit: In other words, let us
assume that the synaptic change and memory are assigned
on the coupling $\epsilon (i)$ at $i$, not on the interaction between
$i$ and $j$.  Thus we take the following model
\begin{equation}
x_{n+1}(i) = (1- \epsilon _n (i))f(x_n (i))+ \frac{\epsilon _n (i)}{N} \sum _j
f
(x_n (j)),
\end{equation}
where $\epsilon _n (i)$ is in-(de-)creased according to inputs on the $i$-th
element.  To be specific, we take the following dynamics; first increase
the coupling to the mean field, according to the input $s_n (i)$ by
\begin{equation}
\epsilon '(i) = \epsilon _{n} (i) +\gamma s_n (i),
\end{equation}
and then rescale the coupling so that the average of $\epsilon (i)$ is
conserved:
\begin{equation}
\epsilon_{n+1}(i) = \epsilon _0 \times \frac{\epsilon '(i)}{\sum_j \epsilon
'(j)}.
\end{equation}
Indeed the latter equation  is introduced as a suppression, and
may be replaced by other forms to suppress the overgrowth
of $\epsilon_{n} (i)$.
What we need here is a mechanism for (a) the suppression of an indefinite
increase of coupling $\epsilon (i)$ and (b) the competition among
elements for the increase of coupling.
When common inputs are applied to
elements $i_0 \leq i \leq i_1$,
synchronization degree of
the oscillations  $x_n (i)$ for $i_0 \leq i \leq i_1$
increases.  Depending on the input strength, two elements $i,j$
with common inputs often synchronize completely.
After inputs are eliminated, a pair of (almost) synchronized elements
remains coherent or correlated.
We have thus achieved clustering according to inputs \cite{Araki}.
This clustering is
often preserved even if the coupling $\epsilon _n (i)$ is again restored to be
homogeneous.
Generally, the clustering is formed according to the structure of inputs.
We have made some simulations with correlated inputs in $p$ groups per
$N/p$ elements.  In particular we take
\begin{equation}
s_n (i)= R_n (int(\frac{i}{p})) + r_n (i)
\end{equation}
where $R_n$ is a random signal with the amplitude 1 but common
for each group $k\frac{N}{p} < i \leq (k+1) \frac{N}{p}$ while $r_n$ is
a noise depending on elements but with much smaller amplitude $\delta $
($\delta  < 1$).  Adopting this form,
inputs are random, but are
strongly correlated within each group of $N/p$ elements.
In Fig.6 we have plotted the spacetime diagram when these inputs are applied.
Correlated motion within each $N/p$  elements is seen.  Some of the
correlation remain even after the inputs are eliminated
( see Fig. 6b)).
To see the correlation in oscillations $\{ x_n (i) \}$ more quantitatively,
we introduce the difference matrix
\begin{equation}
\Delta _{ij} = <(x_n(j)-x_n(i))^2>,
\end{equation}
where $<...>$ is temporal average.  The difference matrix is plotted
in Fig.7 for the inputs in the above structure.  We note that the
partial clustering is formed according to the structure of inputs
as long as the number of input groups is small.  This organization is
not trivial since we have not imposed any direct change of the coupling
to increase the coherence among elements with correlated inputs.
No enhancement of
the connection between pairs is required here.  Indeed, given an element,
the coupling strength takes a same value for any pair between it and
other elements.
The formation of clustering according to inputs works well if
the parameters $a$ and $\epsilon_0$ are chosen so that the number of
clusters there is not far from the expected cluster number by inputs.
If the number of clusters at the corresponding $a$ and $\epsilon_0$ values
is smaller than that of the inputs, some of the groups are fused into
a same cluster.  Relationships between some input groups are
self-organized by the internal dynamics.  Generally,
the above clustering is formed by the balance between internal
dynamics and the external inputs.  In Fig.7 a), for example,
3 out of 4 input groups are mapped into the clustering, but the other group
is not clearly mapped.  In Fig. 7b), two of the four input
groups are mapped, while the other two split into smaller clusters.
\vspace{.1in}
Fig.6 spacetime diagram
\vspace{.1in}
\vspace{.1in}
Fig.7 difference matrix
\vspace{.1in}
There is a recent report of an interesting experiment by Hayashi
\cite{Hayashi}, where synchronization among neurons is
increased when a rat is continuing a task with a feedback to brain.
In our model this corresponds to the growth of $\epsilon (i)$
for all elements.  With further continuing the task, the synchronization
reaches maximum, and the rat's brain goes to
an epileptic state, where the information processing ability is lost.
To avoid such catastrophe, we have to introduce some mechanisms to
suppress the increase of $\epsilon$ in our model.
We may expect that such mechanism exists in the real brain,
and that its breakdown leads to the epilepsy.
\section{Growing Coupled Maps: Origin of Diversity and Differentiation}
One missing feature in our network of chaotic elements so far,
is the possibility to change the degrees of freedom themselves.  In
biological systems,
often the number of elements itself varies.  Such growth of elements
is also seen in the economics.
Here we study a very simple model with the growth of the number of elements,
taking a coupled map model.  A related but more realistic model for
cell division and differentiation is given in
\cite{KKTY}.
We assume that there is a variable $x(i)$ determining the cell state,
and that cells compete with each other for a source term $s$.
Source term $s$ is supplied from outer environments with a constant rate $c$.
The ability to get this source depends on the internal state $x(i)$,
with some nonlinear function $f(x)$.
Thus the dynamics of each $x(i)$ is given by
\begin{equation}
x_{n+1}(i)= x_n (i)+ f(x_n (i)) - S_n;
\end{equation}
\begin{equation}
S_{n} = \frac{c - \sum_j f(x_n (j))}{N}.
\end{equation}
The term $x_{n+1}(i)- x_n (i)= f(x_n (i)) - S_n$ gives a source term
that the element $i$ takes at the time step $n$.  The second condition
assures $\sum_i \{ x_{n+1}(i)- x_n (i) \} =c$, that is,
the sum of the source term balances with that supplied externally.
Further we introduce the following dynamics for cell division and death.
(A) Divide the cell $i$ if $x_n(i)> T_g$; After the division, $x$ of the  cell
$i$ and the new element $N+1$ is assigned to be
 $(x_n(i)- T_g)/2 +\delta$  and $(x_n(i)- T_g)/2 -\delta$, with a
very small random number $\delta$.
(B) Remove the cell $i$ if $x_n(i)< T_d$;
Here we fix $T_g=1$ and $ T_d=0$, although our results are
essentially independent of the choice of $T_g$ and $ T_d$.
The function $f(x)$ is chosen to be $f(x)=\frac{K}{2\pi} sin(2\pi x)$
Time series of the number of cells and the effective number of
clusters are plotted in Fig.8, while
the maximal and average numbers of cells are plotted as
a function of $K$ in Fig.9.  We note that the growth is maximal
around $K \approx 4.3$.
In the corresponding coupled map model with a fixed number of elements
(i.e., globally coupled circle map \cite{GCM-CC}),
the system is in a coherent phase for $K<2$, at the ordered phase
for $2<K<4.1$,  at the partially ordered phase around $4.1<K<4.4$,
and at the turbulent phase for $K>4.4$.
Thus our result suggests that the growth is enhanced at the partially
ordered state.
If oscillations of all elements are synchronized,
they compete for the source term at the same timing.
This hard competition is not good for an effective use of resources.
By the clustering, a sort of time sharing system is constructed.
Thus resources are effectively used with some oredering by elements.
On the other hand, if the elements are completely  desynchronized,
no ordering for the use of resources is possible.
In this case effective use of resources is again impossible.
Thus the growth is enhanced at the partially ordered state,
where synchronization is lost but there remains some ordering.
At the partially ordered state, typical clustering is
quite inhomogeneous.  Some elements form a large cluster,
while there are many other elements with desynchronization.
Cells belonging to a large synchronized cluster
grow much slower than desynchronized cells.
When these synchronized cells divide, the number of cells suddenly
increases by $N_1$, the number of elements in the cluster.
This increase causes a hard competition for resources, and
often leads to spontaneous death of many cells.
Such multiple death reminds us of the programmed death
known in real biological developmental processes.
Our growth and division model may also have relevance to economics, where
many individuals or companies compete for finite resources.
Growth, division, death (bankrupt) factors are important in economics.
Time sharing for resources is useful there.  Economic crash may be
related with the above multiple cell deaths, and may be due to
synchronized behaviors of agents for resources.
The present model, of course, is too simple to the study of
cell growth and differentiation.  Here we have discussed this model
as the simplest illustration.  For a model including metabolic
reaction, and other stimulating results, see \cite{KKTY} in
the present proceeding.  In the present model we have not found a
state corresponding to the stage III in \cite{KKTY}.  Possibly
we need at least a model with two variables, phase and amplitude,
to have the stage III where the separation of poor and rice cells
emerges.
\vspace{.2in}
Fig.8
\section{Summary and discussions}
In the present paper we have discussed relevance of dynamic clustering to
biological problems.
In a network of chaotic elements, they often split into
synchronized clusters due to chaotic instability.
Identical elements spontaneously differentiate.
Thus the clustering can give a basic concept for
the origin of diversity.  Indeed we have
applied the clustering mechanism to cell differentiation
in \S 7.
By using a hypercubic topology and local nonlinear dynamics,
we have studied clusterings with bit structures.
Clusters are spontaneously formed reflecting the bit structure.
Relevant bits are spontaneously formed, which opens up the
possibility of self-organizing genetic algorithms.
Since viruses form quasispecies coded in a hypercubic space
as Eigen et al. discusses  \cite{Eigen},
it may be interesting to search for dynamic clustering
there.
When chaotic instability and averaging by couplings are somewhat balanced,
a partition into clusters is very complex.  In this
partially ordered state, dynamics is also complex with
chaotic itinerancy over ordered states.
Relevance of the partially ordered states and chaotic itinerancy to
biological networks have been discussed throughout the paper.
In the hypercubic topology, chaotic switches of relevant bits
are formed successively.  Besides possible relevance to
dynamics of immune networks and virus populations,
such chaotic itinerancy of bits may be of
use in genetic algorithms.
Relevance of the partially ordered states to ecological and immune
networks is studied with the inclusion of the
change of effective couplings as mutation of mutation rates.
It turns out that the system maintains its stability as
homeochaos, by forming a feedback mechanism to keep the system
around partially ordered states.  The
homeochaos
provides a mechanism of the maintenance of diversity,
important in ecological and othe biological networks.
In a model allowing for the growth of the number of elements,
we have found that effective time sharing system of resources
is formed in partially clustered states.  This study, originally
motivated in the cell differentiation and division,
may be applied to economics, where a breakdown of
the time sharing system by synchronization may lead to
economic crash.
In a GCM with synaptic couplings, we have observed the
clustering formation through some
interference of external inputs and local dynamics.
To have a capacity to construct a map of complex environment,
it is desirable to have a potentiality of complex partitions to
clusters, supported by partially ordered states.
To sum up we have studied classes of extensions of coupled maps
to hypercubic topology, synaptic couplings, growing degrees, and so on,
in order to understand the origin and maintenance of
diversity and complexity in biological networks.
{\sl acknowledgements}
I am grateful to Ichiro Tsuda , Takashi Ikegami, Kenji Araki,
Hatsuo Hayashi, Ad Aertsen, Hiroyuki Ito, Hiroshi Fujii, Walter Freeman,
and Tetsuya Yomo for stimulating discussions.
I am again grateful to
Takashi Ikegami ( for \S3 and \S 4), Kenji Araki
 ( for \S 6), and Tetsuya Yomo (for \S 7),
for exciting discussions through
collaboration on related works \cite{homeo,homeo2,Araki,KKTY}.
The work is partially supported by
Grant-in-Aids for Scientific
Research from the Ministry of Education, Science, and Culture
of Japan.

\pagebreak
Figure Caption
Fig. 1
Snapshot $x(i)$ of hypercubic coupled map (2) with $k=7$ and $a=1.5$,
plotted as a function of $i$, decimal representation of the bit string.
$\epsilon =.4$ for a) and b), $\epsilon=.3$ for c) and d).
a) Two successive time steps (after 100000 steps)
are overlaid for a two-cluster attractor,
split by the condition $******0$ and $******1$.
b) 8 successive time steps are overlaid for a two-cluster attractor,
split by the condition $0******$ and $1******$.
c) 8 successive time steps (after 100000 steps)
are overlaid for a two-cluster attractor,
split by the XOR condition [$***0*0*$ or $***1*1*$] and
[$***0*1*$ or $***1*0*$]. (The attractor is a stable cycle with
period-4).
d) 8 successive time steps are overlaid for a 12-cluster attractor.
(The attractor is chaotic).
\vspace{.1in}

Fig.2: Space-time diagram for the coupled map lattice on a
hypercubic lattice (2) with $a=1.54$, $\epsilon =.3$
and $k=6$ (i.e., $N=2^6$).
On the corresponding pixel at a given time and element,
a bar with a length
proportional to $(x_n(i,j)-0.1)$ is painted if $x_n(i,j)>.1$.
(a) Every  4th time step is plotted from 50000 to 52400.
(b) Continued from a), but for the time steps 60000 to 62400.
\vspace{.1in}

Fig. 3
Snapshot plots of $h^s _n (i)$ for the model with
mutation of mutation rates described in the text.
$a=3.5$, $\beta=7.0$, $k=7$ and $j_{max}=30$.
(a) time step 9000 (b) 18000 (c) 21000 (d) 24000.
\vspace{.1in}

Fig.4
(a) Time series of the effective degrees of freedom, corresponding
to the simulation of Fig.3.  An effective cluster is
defined  as follows; if $x(i)$ and $x(j)$ agree within the precision
$2^{-13}$. Successive drops of the degrees are seen, which
corresponding to a bit-clustered state like Fig.3a).
(b) Corresponding temporal evolution of the average mutation level $j$.
\vspace{.1in}
Fig.5  Snapshot of the globally coupled map (5), with $a=1.6$,
$N=10^5$. $x_n (i)$ at the time step $n=5000$
is plotted as a function of $i$.  The coupling $\epsilon(i)$ is
ordered as $\epsilon (i)= 0.4 (i/N)$.
(b) Successive snapshots $x_n (i)$
at the time steps 5001, 5002, 5003, and 5004 are overlaid.
\vspace{.1in}
Fig.6
Space-time diagram for the globally coupled map eq. (6) with
eqs. (7) and (8) with the inputs eq.(10),
with $p=4$.  $a=1.85$, $\delta=0.1$, $\gamma =0.1$, $N=128$.
On the corresponding pixel at a given time and element,
a bar with a length
proportional to $(x_n(i,j)-0.1)$ is painted if $x_n (i,j)>.1$.
(a) Every  16th time step is plotted from 8000 to 16000.
(b) Continued from a), after the inputs are stopped at the time step 20000.
Over the time steps 24000 to 32000.
\vspace{.1in}
Fig.7
Difference matrix eq. (10) averaged over 20000 steps, after
inputs of eq. (9) with $p=4$ and $\delta =0.1$ are applied for
20000 steps.
The synaptic dynamics (7) and (8) are used with $\epsilon _0 =0.3$,
and $\gamma =0.1$
(a) $N=16$, and $a=1.85$
(b) $N=32$, and $a=1.75$.
\vspace{.1in}
Fig.8
Temporal evolution of the number of cells $N$ (thick line)
and the effective cluster number ( dotted line).
The effective cluster number is defined as that of clusters
with the precision $10^{-4}$ ( checking
the equality between $x(i)$ and $x(j)$ with this precision).
Simulations are carried out with the use of eq. (11) with
$c=0.1$, starting from one cell ($N=1$).
(a) $K=2.4$ (b) $K=3.35$ (c) $K=3.45$
\vspace{.1in}
Fig.9
The average ( thick line) and maximal number of cells
over the time steps 5000  to 15000.
Simulations are carried out with $c=0.1$,
and  starting from one cell.

\end{document}